\documentclass[useAMS,usenatbib,a4]{mn2e}
\newcommand{\thickhline}{\noalign{\hrule height 1.0pt}}
\usepackage{graphicx} % This allows you to call \includegraphics
\usepackage{amsmath} % Allows certain special characters and symbols
\usepackage{amssymb} % Allows certain special characters and symbols
\usepackage{rotating}   % to rotate the expressions and tables.
\usepackage{multirow}
\def\etal{{\it et al}}
\def\deg{^{\circ}}
\def\P3hat{{\mathaccent 94 P}_3}

\def\eg{{\it e.g.}}

\def\aap{A\&A}
\def\aaps{A\&A Suppl.}
\def\apj{Ap.J.}
\def\apjl{Ap.J. Lett.}
\def\apjs{Ap.J. Suppl.}
\def\jaa{{\it J. Astr.\&Astron.}}
\def\mnras{M.N.R.A.S.}

\def\chaaps{{\it Ch. J. Astr.\&Astrop.}}
\newcount\notenumber
\def\clearnotenumber{\notenumber=0}
\def\note{\advance\notenumber by1 \footnote{$^{\the\notenumber}$}}
\clearnotenumber
\title[B0943+10 VI: The `Q'-mode precursor and comparison with B1822--09]{The beam topology and dynamic emission properties of pulsar B0943+10 --- VI.  
Discovery of a `Q'-mode precursor and comparison with pulsar B1822--09}

\author[Isaac Backus, Dipanjan Mitra, \& Joanna Rankin] 
{Isaac Backus$^{1}$, Dipanjan Mitra$^{2}$ \& Joanna M. Rankin$^{3,1}$ \\ 
$^1$Physics Department, University of Vermont, Burlington, VT 05405 USA\thanks{Isaac.Backus@uvm.edu; Joanna.Rankin@uvm.edu} \\
$^2$National Centre for Radio Astrophysics, Ganeshkhind, Pune 411 007 India\thanks{dmitra@ncra.tifr.res.in; } \\
$^3$Sterrenkundig Instituut `Anton Pannekoek', University of Amsterdam, NL-1098 SJ \\}

\date{Unreleased}

\pagerange{\pageref{firstpage}--\pageref{lastpage}} \pubyear{2004}

\def\LaTeX{L\kern-.36em\raise.3ex\hbox{a}\kern-.15em
    T\kern-.1667em\lower.7ex\hbox{E}\kern-.125emX}

\begin{document}

\label{firstpage}

\maketitle

\begin{abstract}
This paper reports new observations of pulsars B0943+10 and B1822--09 
carried out with the Arecibo Observatory (AO) and the Giant Metrewave Radio 
Telescope (GMRT), respectively.  Both stars exhibit two stable emission modes.  
We report the discovery in B0943+10 of a highly linearly polarized ``precursor'' 
component that occurs primarily in only one mode.  This emission feature closely 
resembles B1822--09's precursor which also occurs brightly in only one mode.  
B0943+10's other mode is well known for its highly regular drifting subpulses 
that are apparently produced by a rotating ``carousel'' system of 20 `beamlets.'  
Similary, B1822--09 exhibits subpulse-modulation behavior only in the mode 
where its precursor is absent.   We survey our 18 hours of B0943+10 observations 
and find that the `sideband'-modulation features, from which the carousel-rotation 
time can be directly determined, occur rarely---less than 5\% of the time---but 
always indicating 20 `beamlets'.  We present an analysis of B1822--09's 
modal modulation characteristics at 325-MHz and compare them in detail with 
B0943+10.  The pulsar never seems to null, and we find a 43-rotation-period 
$P_3$ feature in the star's `Q' mode that modulates the interpulse as well as 
the conal features in the main pulse.  We conclude that B1822--09 must have 
a nearly orthogonal geometry and that its carousel circulation time is long 
compared to the modal sub-sequences available in our observations, and
 the mainpulse/interpulse separation is almost exactly 180\degr.  We conclude 
 the precursors for both stars are incompatible with core-cone emission.  
 We assess the interesting suggestion by Dyks \etal\ that downward-going 
radiation produces B1822--09's precursor emission.  
\end{abstract}


\section*{I. Introduction} 
Among the well investigated pulsars that exhibit the phenomenon of ``mode 
switching'', B0943+10 provides one of the clearest examples of two discrete 
modes, both exhibiting distinct, fully characterizable behaviors (Suleymanova 
\& Izvekova 1984).  In this paper, the sixth in a series describing B0943+10 
analyses, we (somewhat abashedly) report a newly discovered precursor 
feature in the profile of this intriguing star which is bright in the `Q'-mode and 
nearly undetectable in the `B'-mode, suggesting a strong similarity to another 
well studied pulsar with highly discrete modes, B1822--09 (or J1825--0935).   

During its weak, `Q'uiescent mode, B0943+10 is well known to exhibit a chaotic 
subpulse-modulation behavior (Suleymanova \etal\ 1988).  By constrast, its 
`B'urst mode has been of great interest throughout this series, in that it now 
stands as the paradigm example of regularly drifting subpulses.  Analyses 
have repeatedly shown that the observed `B'-mode pattern of subpulses 
results from a rotating carousel of just 20 `beamlets,' such that consistent 
subbeam-carousel maps have been constructed for observations at a number 
of different epochs and frequencies (Deshpande \& Rankin 2001; Asgekar \& 
Deshpande 2001; hereafter Paper I and Paper II of this series).  

According to the polar cap emission theory of Ruderman \& Sutherland (1975), 
the observed subbeam carousel is thought to result from ``spark''-induced 
columns of relativistic primary plasma directed into the `open' polar flux tube and 
precessing around the magnetic axis under the action of ${\bf E}\times$${\bf B}$ 
drift.

An important finding for developing a fuller model of the polar cap emission 
region of B0943+10 was the presence of evenly spaced `sidebands' surrounding 
the primary modulation feature associated with drifting subbeams.  Paper I argued 
that these sidebands are the signatures of a modulation on the true (un-aliased) 
primary drift-modulation frequency ($f_3$) corresponding to the circulation time 
of a carousel of 20 beamlets ($\P3hat = n P_3$, where $n$ is the number of 
subbeams and $P_3 = f^{-1}_3$), and evidence presented in Paper II corroborated 
this conclusion.  However, an important consideration in understanding the 
physical origin of the observed fluctuation spectra was left unresolved: how often 
and under what circumstances do the sidebands appear?

In this paper we continue our analysis of B0943+10 and compare its behavior to 
that of another famous mode-switching pulsar B1822--09.  Several basic similarities 
between the stars prompt further investigation: both have comparable periods 
(1.098 s for B0943+10; 0.769 s for B1822--09); and both have estimated surface 
magnetic fields on the order of $10^{12}$ G (ATNF Pulsar Database\footnote{http://www.atnf.csiro.au/research/pulsar/psrcat}).  Furthermore, 
both are detectable as X-ray emitters (Zhang, Sanwal \& Pavlov, 2005; Alpar, 
Guseinov, Kiziloglu, \& \"Oegelman 1995).  

More important are the remarkable modal similarities between these two pulsars. 
B1822--09 has fascinated researchers because of its several discrete behaviors.  
Like B0943+10, it switches between two modes (Fowler \etal\ 1981; Gil \etal\ 1994): 
its so called `Q'uiescent mode, which exhibits regularly drifting subpulses (as in the 
`B' mode of B0943+10) and an interpulse; and its `B'urst mode in which a bright 
precursor component `turns on' after being only faintly detectable, the regular 
modulation disappears (as in the 
`Q' mode of B0943+10)\footnote{In this paper we choose to continue the terminology 
established by Fowler \etal\ for the stars' two modes.  The `B' mode of B0943+10 
exhibits behaviors such as drifting subpulses similar to the `Q' mode of B1822--09, 
while the `Q' mode of B0943+10 and the `B' mode of B1822--09 both display a 
precursor (see Table~\ref{tab:modes})  The names derive from the relative intensity 
of the two modes and do not represent their most physically significant properties.  
In short, the `B' mode in one star does not correspond to the `B' mode of the other.}, 
and the interpulse turns off.  Nonetheless, the anti-correlation between B1822--09's 
interpulse and precursor (hereafter IP and PC; Fowler \& Wright 1982) presents a 
difficulty for current emission models:  if B1822--09 is a nearly orthogonal rotator, 
as has been argued, how does information transfer from one magnetic pole to the 
other?  It has even been argued that the IP and PC originate from the same emission 
region which reverses emission direction in its two modes (Dyks, Zhang \& Gil 2005).

The organization of the paper is as follows.  In \S II we describe the observations 
of B0943+10 and B1822--09 used in this study.  In \S III we present an analysis of 
18 hours of B0943+10 `B' mode observations and describe the discovery of 
two new instances of sidebands.   Then, in \S IV we report the discovery of a 
`precursor' component in B0943+10.  

In \S V we report a study of two GMRT 325-MHz observations of B1822--09, 
the first ever detailed analysis at meter wavelengths.  Fluctuation-spectral 
evidence is reported suggesting that the main pulse (hereafter MP) and IP 
are linked, contrary to the Dyks \etal\ reversal model.  We find that B1822--09's 
modal behaviors, profile forms, and the polarization properties of its MP and PC 
are comparable to those of B0943+10.  And like B0943+10, B1822--09 never 
seems to null.  We further argue that B1822--09 is indeed an orthogonal rotator.  
In \S VI we discuss the implications of our 
findings for the emission reversal model of Dyks \etal\ and for a non-radial oscillation 
model proposed by Clemens \& Rosen (2004) to explain the observed supbulse-drift 
of B0943+10.  Finally, in \S VII, we review our findings, and the case available for assimilating the 
properties of B1822--09 and B0943+10.  In particular, the PC emission in the 
two stars appears to be neither of the conal nor core type, and its geometry could 
associate it with the so-called ``outer gap'' where the high-energy emission from 
pulsars is thought to originate.

\section*{II. Observations} 
\label{sec2:II} 
The B0943+10 observations used in our analyses were made using the 305-m 
Arecibo Telescope in Puerto Rico (hereafter AO). The 327-MHz (P band) polarized 
pulse sequences (hereafter: PS) were acquired using the upgraded instrument 
together with the Wideband 
Arecibo Pulsar Processor (WAPP\footnote{http://www.naic.edu/\textasciitilde 
wapp}) on a number of different days over a four-year period as detailed in 
Table~\ref{tab:observations}.   
\begin{table}
 \begin{center}
   \caption{B0943+10 and B1822--09 Observations} \label{tab:observations}
   \begin{tabular}{cccc}
   \hline
     \textbf{MJD}  & \textbf{Frequency} & \textbf{Resolution} & \textbf{Length}  \\
          & (MHz) & ($\deg$longitude) & (in pulses) \\
   \thickhline       
	\multicolumn{4}{c}{\textbf{AO B0943+10}} \\
   \thickhline
	$^a$48914 & 430 & 0.330 & 986 \\
	$^b$52709 & 327 & 0.352 & 6748 \\
	$^b$52711 & 327 & 0.352 & 6809 \\
	52832 & 327 & 0.352 & 7559 \\
	52840 & 327 & 0.352 & 7275 \\
	52916 & 327 & 0.352 & 6560 \\
	52917 & 327 & 0.352 & 3024 \\
	53491 & 327 & 0.459 & 5841 \\
	53492 & 327 & 0.459 & 6825 \\
	53862 & 327 & 0.352 & 5041 \\
	54016 & 327 & 0.352 & 6012 \\
	54630 & 327 & 0.352 & 7569 \\
	54632 & 327 & 0.352 & 6656 \\
 	\thickhline
 	\multicolumn{4}{c}{\textbf{GMRT  B1822--09\textbf{}}}  \\
  	\thickhline	
 	53780 & 325 & 0.240 & $^c$2077 \\
	$^b$54864 & 325 & 0.240 & 2106 \\
 	 \hline
 \end{tabular}
 \end{center}
 \tiny
 The files of the B0943+10 observations used in this analysis were resampled 
 from their original resolution.  Higher resolution is available, but was unnecessary 
 for the analysis presented here. \\
   $^a$Only 41$\deg$ are available for this observation \\
   $^b$These observations lack polarimetry \\
   $^c$The original observation contains 2300 pulses.  Due to interference, we 
   ignore the last 223 pulses here.
\end{table}
The auto- and cross-correlations of the channel 
voltages were three-level sampled and produced by receivers connected to 
linearly (circularly during the MJD interval 53289 to 54629) polarized feeds. 
Upon Fourier transforming, sufficient channels were synthesized across a 
25-MHz (50-MHz after MJD 54630) bandpass, providing resolutions of about 
1 milliperiod of longitude. The Stokes parameters have been corrected for 
dispersion, interstellar Faraday rotation, and various instrumental polarization 
effects.  Some of the PSs have been discussed in previous papers in this series; 
however this paper presents the 6 days of 2+ hour observations since MJD 53491.  

The two observations of B1822--09 were carried out using the Giant Meterwave 
Radio Telescope (hereafter GMRT) near Pune, India, using the same techniques as 
described in Mitra, Rankin \& Gupta (2007).

Table~\ref{tab:modes} 
\begin{table}
 \begin{center}
   \caption{Mode Changes in B0943+10 and B1822--09 Observations} \label{tab:modes}
   \begin{tabular}{cccc}
   \hline
     \textbf{MJD}  & \textbf{Modes} & \textbf{$^a$Switch} & \textbf{Length}  \\
          &       & (pulse) & (in pulses) \\
   \thickhline       
	\multicolumn{4}{c}{\textbf{AO B0943+10}} \\
   \thickhline
	48914* & B to Q & 816 & 986 \\
	52709* & Q to B & 2540 & 6748 \\
	52711 & B & -- & 6809 \\
	52832 & Q to B & 5266 & 7559 \\
	52840 & B & -- & 7275 \\
	52916 & Q to B & 1755 & 6560 \\
	52917 & B & -- & 3024 \\
	53491 & B & -- & 5841 \\
	53492 & B & -- & 6825 \\
	53862* & B & -- & 5041 \\
	54016 & Q & -- & 6012 \\
	54630 & B & -- & 7569 \\
	54632 & Q to B & $^b$3100 & 6656 \\
 	\thickhline
 	\multicolumn{4}{c}{\textbf{GMRT  B1822--09}}  \\
  	\thickhline	
 	\multirow{6}{*}{53780} & B to Q & 200 & \multirow{6}{*}{2077} \\
 	 & Q to B & 770 & \\
 	 & B to Q & 1095 & \\
 	 & Q to B & 1200 & \\
 	 & B to Q & 1475 & \\
 	 & Q to B & 1755 & \\
	  &  &  &  \\
	54864 & Q & -- & 2106 \\
 	 \hline
 \end{tabular}
 \end{center}
 \tiny
 * Sidebands surrounding the primary modulation feature are observable in the fluctuation spectra of these observations (See \S III). \\
  $^a$Because the mode changes of B1822--09 occur over several pulses, these values represent an approximate boundary of the switch. \\
  $^b$This represents an approximation because of interference at the mode switch boundary.
\end{table}
outlines the occurrence of different emission modes present in our observations.  The spectral analysis techniques utilized in this paper were first presented and 
explained in detail in Paper I.  We would ask the reader to refer to that paper for a 
complete description.

\section*{III. Sidebands in B0943+10}

\begin{figure}
\begin{center}
\includegraphics[width=77mm]{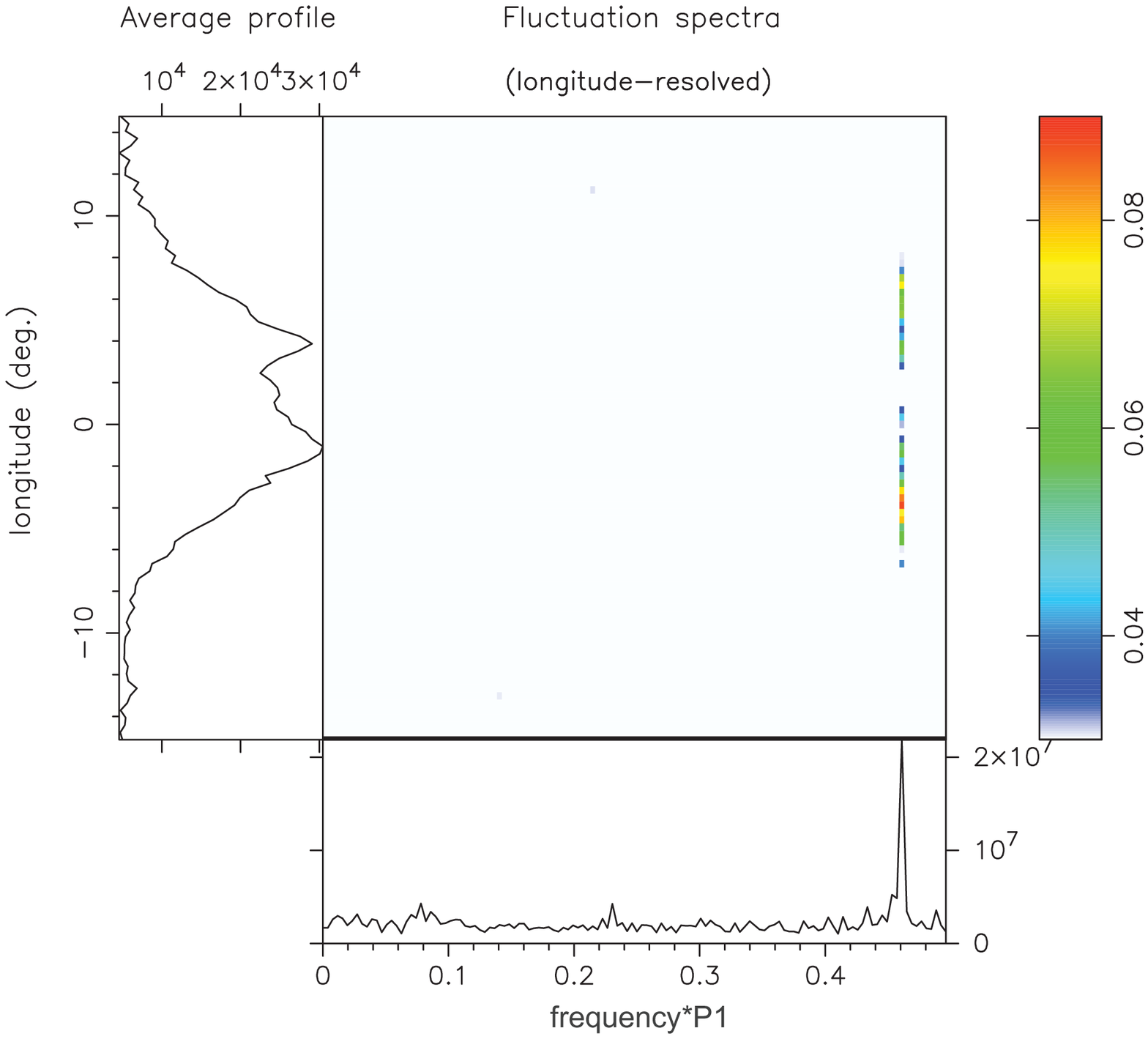}
\caption{`B' mode Longitude-resolved fluctuation (hereafter LRF) spectra for the MP of 
B0943+10 at 430 MHz, averaged over pulses 106-361 of MJD 48914 using 
a 256-point FFT.  The average profile is given at the left of the figure and the 
integral spectrum is at the bottom.  The central panel shows the amplitude 
of the features.  This is the first known instance of sidebands surrounding the 
primary modulation feature of B0943+10.  It was studied at length in Paper I.  
At about 0.026 cycle/$P_{1}$, the sideband spacing represents an harmonic 
relationship with the first-order alias of the large primary modulation feature, 
providing evidence that the pattern of drifting subpulses in B0943+10 comes 
from a rotating carousel of 20 ``sparks'' of bright emission.  The intensity scale 
is arbitrary.} 
\label{fig:sidebands1}
\end{center}
\end{figure}

An important finding of the first paper in this series was the presence of sidebands 
surrounding the primary subpulse drift-modulation feature.  Both the sidebands 
and the primary feature arise only during the `B' mode.  As argued in Paper I, these sidebands, shown 
in the longitude-resolved fluctuation (hereafter LRF) spectra of Figure~\ref{fig:sidebands1}, 
represent a ``tertiary'' modulation of the phase-modulated ``drift'' feature.  
Non-uniformity within a regular pattern generates amplitude modulation.  Thus, unless 
all the subbeams are perfectly identical in their amplitude and spacing---or are 
totally random---we would expect to detect such a tertiary periodicity corresponding 
to the rotation period (or circulation time) of the entire carousel ($\P3hat$).  Using 
the 430-MHz observation, Paper I determined these sidebands to fall symmetrically 
at 0.02680 $\pm$ 0.00037 cycle $P_{1}^{-1}$ above and below the primary feature 
at 0.535 cycle $P_{1}^{-1}$.  
That they are so symmetric and narrow indicates a regular amplitude 
modulation (of the phase modulation).

\begin{figure}
\begin{center}
\includegraphics[width=77mm]{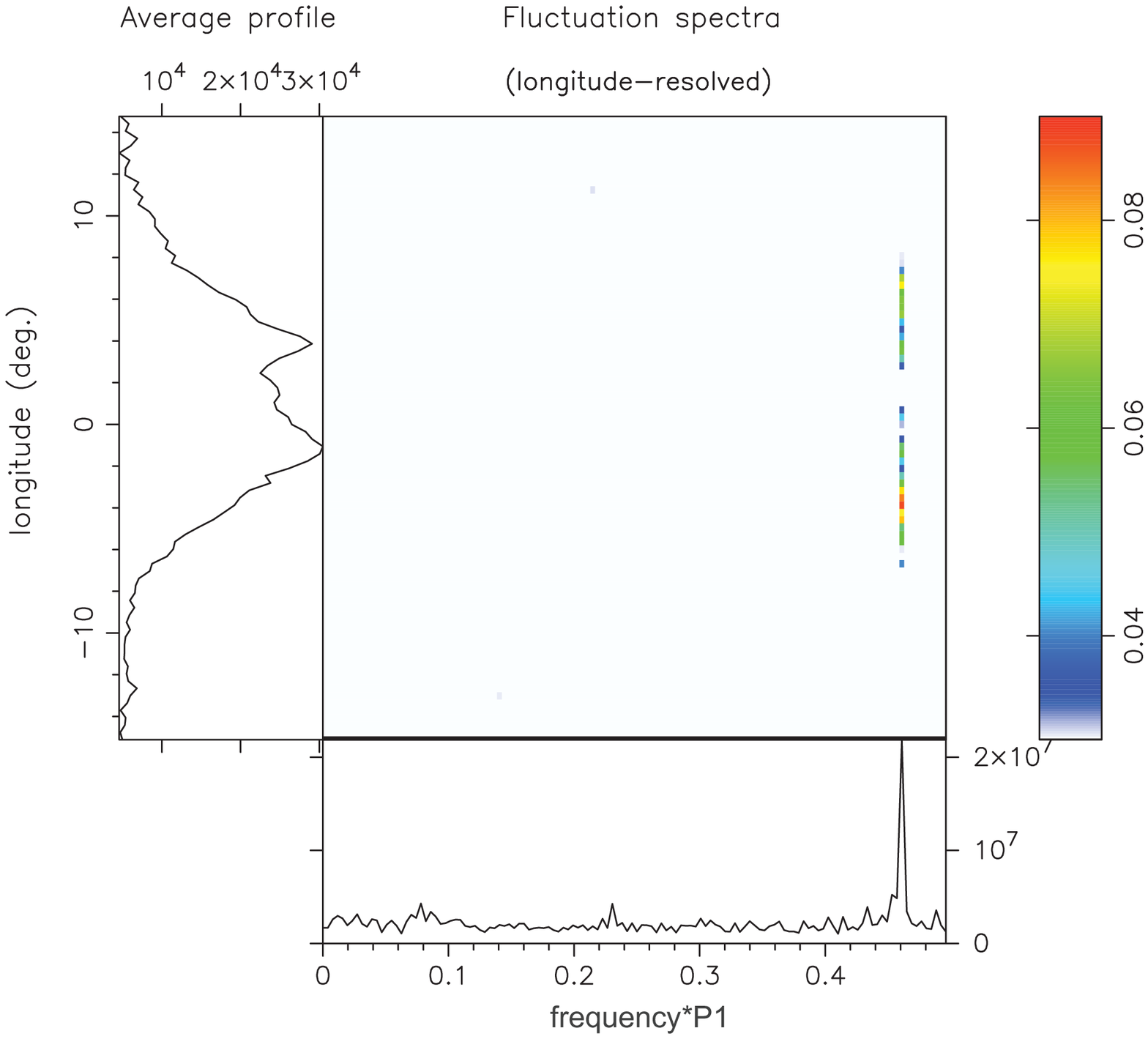}
\caption{`B' mode LRF spectra (as in Fig.~\ref{fig:sidebands1}) for the MP of B0943+10 at 
327 MHz, averaged over pulses 4085-4340 of the MJD 52709 PS using a 
256-point FFT.  Weak sidebands are present, surrounding the main feature; 
their remarkably even spacing and their persistence over several hundred 
pulses allows us to conclude that they represent a physically significant 
modulation of the primary feature at some 0.46 cycle $P_{1}^{-1}$.  Note 
also that the primary feature strongly modulates the two subcomponents, 
while the center of the profile is much less modulated.} 
\label{fig:sidebands2}
\end{center}
\end{figure}

\begin{figure}
\begin{center}
\includegraphics[width=77mm.]{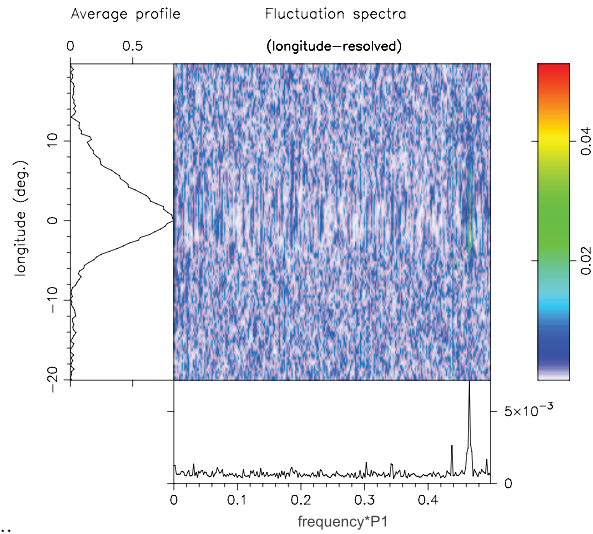}
\caption{`B' mode LRF spectra (as in Fig.~\ref{fig:sidebands1}) for B0943+10 at 
327 MHz, averaged over pulses 260-771 of the MJD 53862 PS using a 
512-point FFT.  The stability of the sidebands in this observation allowed us 
to measure their spacing with high precision.  We were able to average 
over 512 pulses without washing out the modulation, allowing the use of a 512-point 
FFT.  Whereas before the sidebands were symmetric, in this observation one 
is clearly `taller' than the other.} 
\label{fig:sidebands3}
\end{center}
\end{figure}

The brevity of the 430-MHz observation analyzed in Paper I prevented the 
determination of how often these sidebands arise.  We are now able to report 
the results of an analysis based on a wealth of observations, and we find that 
the sidebands are rarely present in B0943+10.  They are in fact only known to 
occur on three separate occasions and of course in the B mode: on MJD 48914 in Paper I 
at 430-MHz (see Fig.~\ref{fig:sidebands1}); and in the 327-MHz observations on 
MJD 52709 and MJD 53862 (see Figs.~\ref{fig:sidebands2}~and~\ref{fig:sidebands3}).  
Out of some 58,000 `B' mode pulses now available in the AO PSs---comprising 
18 hours of observations---sidebands can be discerned in fewer than 
3,000.  When they do appear, the sidebands are stable for several hundreds of 
pulses---which indicates that this tertiary modulation can persist over many 
times the 37-$P_1$ carousel-circulation time---and yet they vanish for many 
hours at a time.  They never seem to persist for more than about 18 mins.

We can conclusively corroborate several of the findings of Paper I.  The sidebands 
never appear to be accompanied by any other pairs, nor is there evidence of 
any other tertiary modulation of the primary feature in PSs where 
the sidebands are not present.  The pair of modulation features are always 
remarkably evenly spaced, the difference in their spacings from the primary 
feature always being less than 3\% of their actual spacing.  The inverse of 
this spacing remains commensurate with the carousel circulation time 
calculated as 20 $P_3$, though the agreement is strongest in the 430-MHz 
observation.  

In the MJD 52709 observation (see Fig.~\ref{fig:sidebands2}), the sidebands 
occur during a roughly 550-pulse interval, during which $\P3hat$ determined 
from the alias of the primary modulation feature is 37.008$\pm$0.013 $P_1$.  
In agreement with the findings of Paper I, these sidebands are of nearly identical 
height, implying an amplitude modulation.  In the MJD 53862 observation, 
sidebands are detectable for some 1000 pulses (see Fig.~\ref{fig:sidebands3}).  
$\P3hat$, measured from the primary modulation feature, is 37.376$\pm$0.005.  
One interesting difference is present in this observation: the sidebands are 
significantly asymmetric; the `right' sideband is only 63\% the height of the `left' 
one.  This indicates a mixture of amplitude and phase modulation (see Paper I).

\begin{figure}
\begin{center}
\includegraphics[width=60mm,angle=-90]{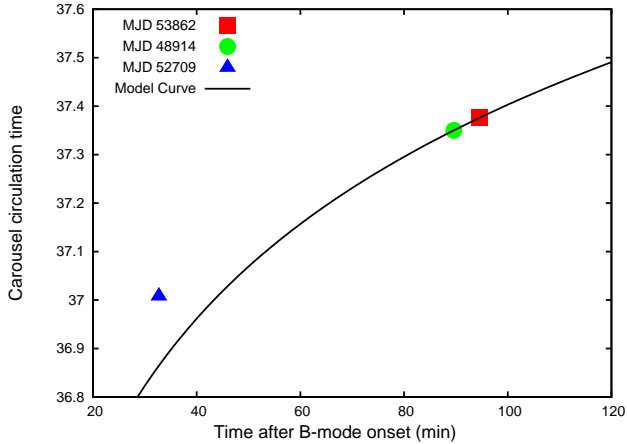}
\caption{$\P3hat$ {\it vs.} the time after `B'-mode onset of the three known 
occurences of sidebands in B0943+10, along with a model 
curve representing the relationship established in Paper~IV: $t = 1.826 \times 
10^{-32} \exp[2.077 \P3hat]$.  There is no discernible relation between the 
sideband occurrence and the evolution of the `B' mode.  The time positions 
of the MJD-48914 and 53862 observations are calculated from $\P3hat$, 
and should be considered estimations; signficant deviation from the model 
curve is expected, but not plotted here. } 
\label{fig:p3hat_vs_time}
\end{center}
\end{figure}

Aside from the rarity of the sidebands, a significant finding is that their 
appearance seems in no way correlated with the evolution of the `B' mode.  
In the 430-MHz observation, the sidebands appear at the end of a `B' mode 
episode, immediately before the transition to the `Q' mode.  In the MJD 52709 
observation, sidebands appear only about 28 minutes after `B' mode onset.  
Then, sidebands appear for 18 minutes at the beginning of the MJD 53862 
observation, and disappear for its remaining 74 minutes (all in `B' mode).  
Using two relationships established in Rankin \& Suleymanova (2006; 
hereafter Paper IV), we can estimate how long after `B' mode onset the 
beginning of this observation lies.  (a) $t = -\tau \ln[(A(2/1)-0.17)/1.16] =$ 
$\sim$100 min, where A(2/1) is the amplitude ratio of the two components 
comprising the MP, and $\tau$ is the characteristic time of some 73 min.  
(b) $t = 1.826 \times 10^{-32} \exp[2.077 \P3hat] =$ $\sim$95 min.  These 
computations are only approximate, but we can conclude that in the MJD 
53862 observation, the sidebands show up around an hour further into the 
`B' mode than they do on MJD 52709 (see Fig.~\ref{fig:p3hat_vs_time}).

\begin{figure}
\begin{center}
{\large {\bf `Q' mode}}
\includegraphics[width=80mm,angle=-90.]{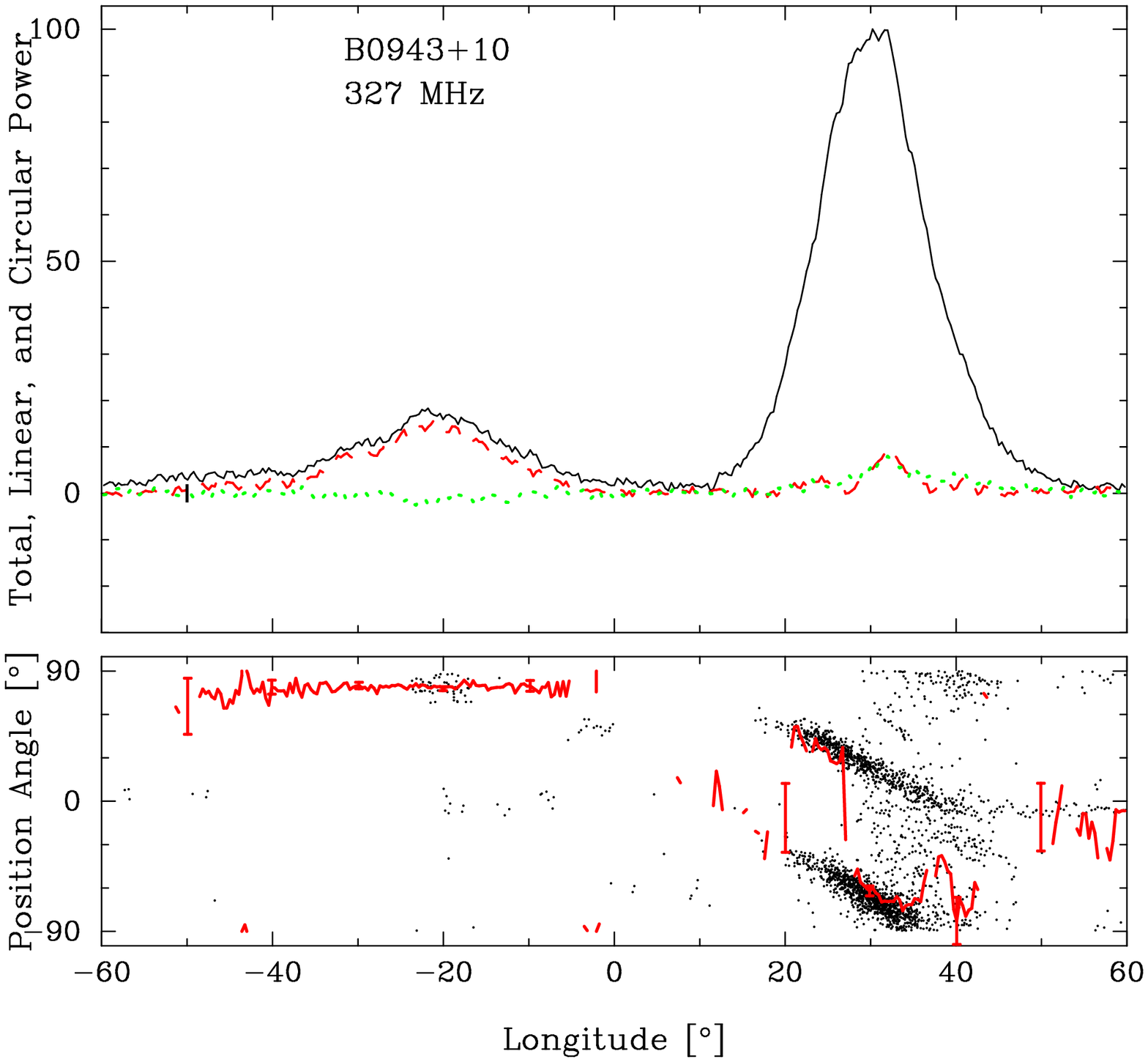}
{\large {\bf `B' mode}}
\includegraphics[width=78mm,angle=-90.]{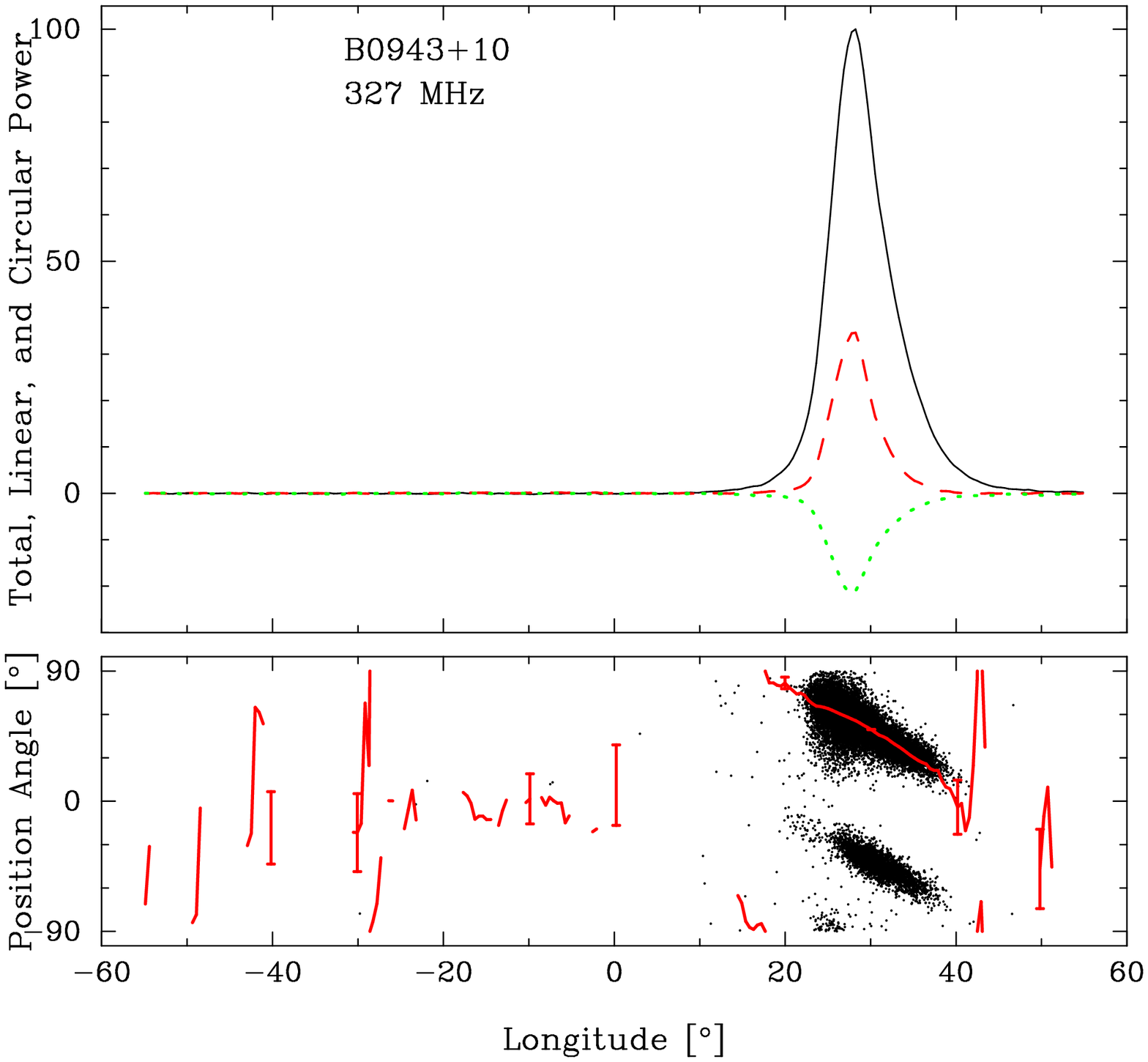}
\caption{Polarization profiles and PPA histograms for B0943+10's `Q' (1050 
pulses of MJD 52832) and `B' (4000 pulses from MJD 53492) modes, respectively.  
{\bf `Q' mode (top)}:  The PC is highly linearly polarized, with almost no circular 
polarization, and note the unusual flat polarization position angle (PPA) traverse; whereas the MP is almost 
completely depolarized by nearly equal levels of OPM power (visible as parallel 
``tracks'' in the PPA distribution, separated by 90$\deg$).  {\bf `B' mode (bottom)}:
Here the MP retains significant primary polarization-mode power, and its PPA 
traverse is well defined.  The PPA regularity around --20\degr, which is only seen 
in very long integrations, may represent weak secondary polarization-mode 
(hereafter SPM) PC power in its `off' state.  The upper and lower panels display 
the total power (Stokes $I$), total linear polarization ($L$ [=$\sqrt{Q^2+U^2}$]; 
dashed red) and circular polarization ($V$ [LH-RH]; dotted green) (upper), and 
the polarization angle ({\it PPA} [=${1\over 2}\tan^{-1}(U/Q)$]) (lower). Individual 
samples that exceed an appropriate $>$2 $\sigma$ threshold (derived from 
off-pulse $L$) appear as dots with 
the average PPA (red curve) overplotted.  The PPAs are approximately absolute 
(see text).  The intensity scale is arbitrary.} 
\label{fig:B0943_mode_polarization}
\end{center}
\end{figure}

\section*{IV. Precursor Discovery in B0943+10}
We now introduce the newly discovered presence of a `precursor' component 
in B0943+10 which occurs strongly only in the `Q' mode.  Measuring from the 
center of the half-power point, the PC lies 52$\deg$ before the MP, as can be 
seen in the upper panel of Figure~\ref{fig:B0943_mode_polarization}.  Because 
previous analyses of B0943+10 have focused almost exclusively on its `B'-mode 
characteristics, most ``working'' PSs were restricted for convenience to a 40-60$\deg$ 
window surrounding the MP; and when not the occasionally present emission 
in the 40$\deg$-longitude range of the PC was at first dismissed as interference.  

During the `B' mode, the PC emission levels are comparable to the noise level, 
resulting in an integrated profile in which the precursor appears absent (see the 
lower plot of Fig.~\ref{fig:B0943_mode_polarization}).  During a `Q'-mode interval, 
the PC is $\sim$18\% of the intensity of the MP (which is itself both weaker and 
broader than in the `B' mode).  Though weaker, the PC is actually some 1.7 
times wider than the MP at half power ($\sim$25$\deg$ and $\sim$15$\deg$, 
respectively).  The PC switches off immediately at `B' mode onset, producing no integrated 
emission above the noise level.  We have no full-longitude observations of the 
`B'-to-`Q' mode transition, so the behavior of the PC at this boundary is unknown.
 
\begin{figure} 
\begin{center}
\includegraphics[width=84mm]{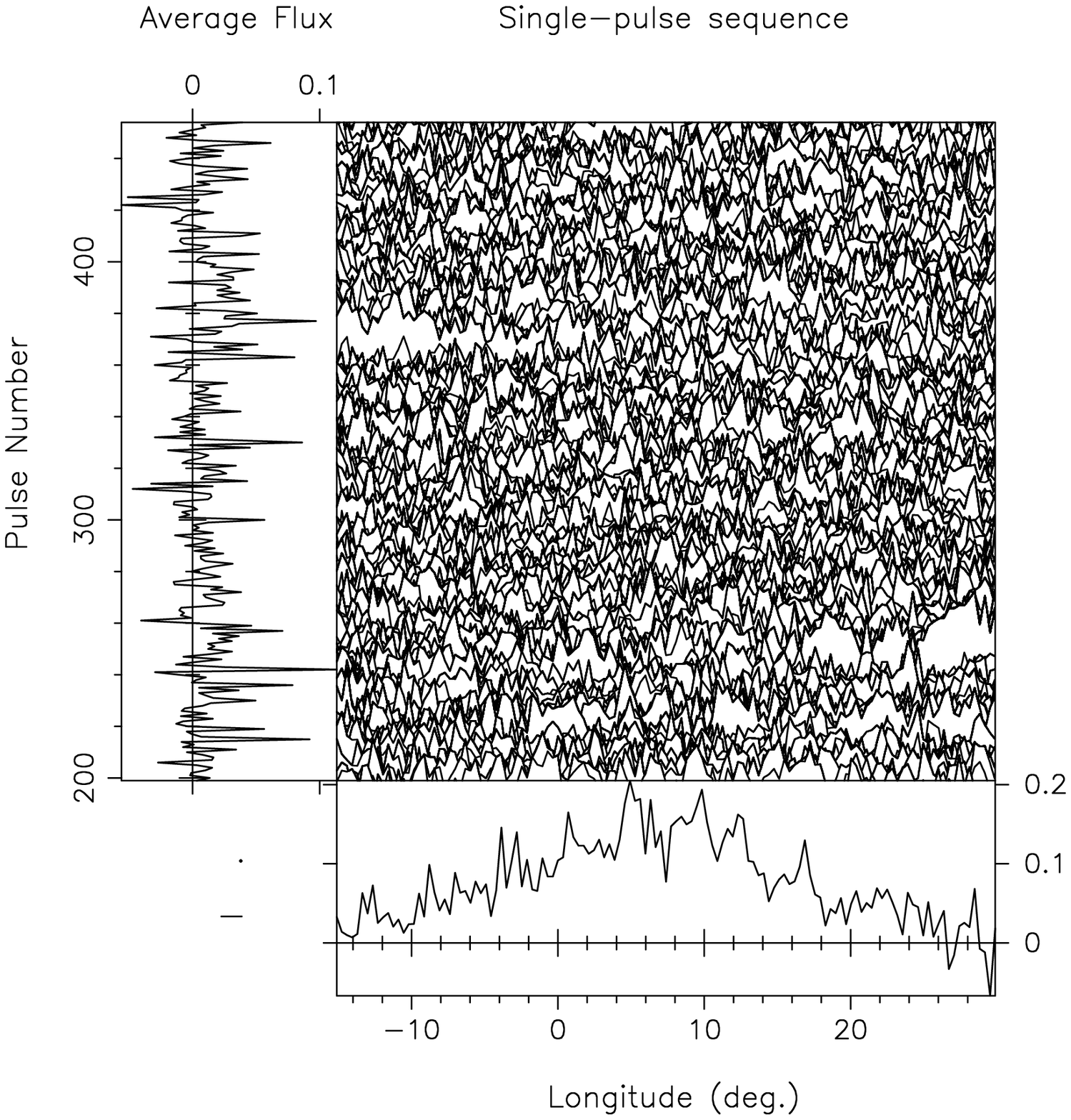}
\includegraphics[width=84mm]{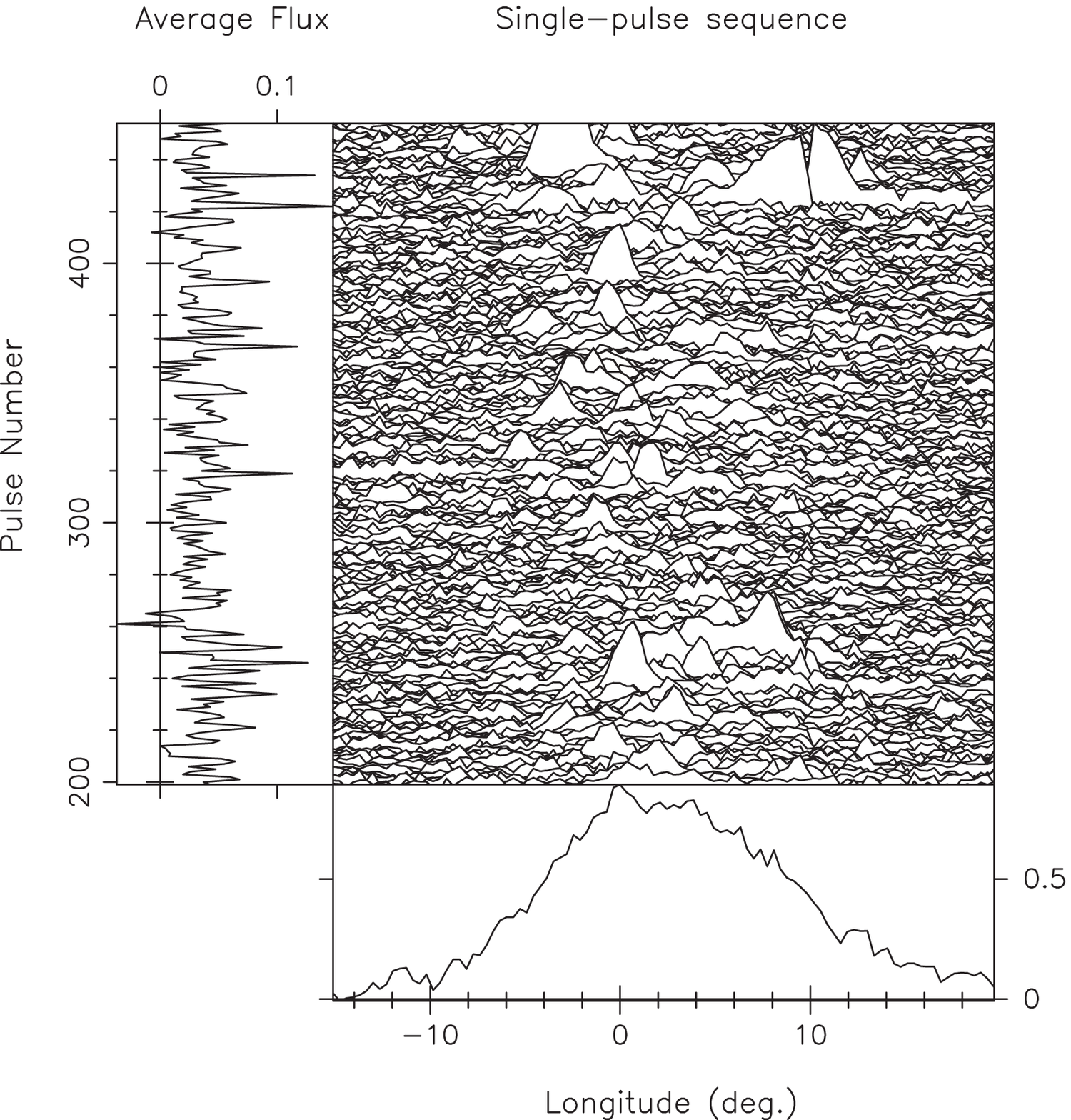}
\caption{A 256-pulse sequence, taken from the MJD 52832 observation. 
The precursor (top) and main pulse (bottom) in the `Q' mode of B0943+10.  
The individual PC pulses are composed of many small spikes of emission, 
while the individual pulses of the MP are comprised of fewer, broader 
subpulses.  The intensity scale is arbitrary.} 
\label{fig:spiky_emission}
\end{center}
\vspace{100mm}
\end{figure}

While the PC and the MP are regulated by the same modes, their properties and 
behaviors are otherwise distinct.  During the `Q' mode, when the PC is most 
prominent, its emission is nevertheless sporadic.  Individual pulses are composed 
of many short spikes of emission, as shown in Figure~\ref{fig:spiky_emission}, 
and are typically difficult to distinguish from the noise.  It is possible that the 
PC and the `Q'-mode MP null, but the sporadic pulse shapes and low intensity 
of the PC make analysis of individual pulses difficult in our observations: nulls 
simply cannot be distinguished from noise fluctuations.
As was pointed out prominently in Paper I, the MP is itself more sporadic in the 
`Q' mode than in the `B' mode, but it is still comprised of recognizable subpulses 
as opposed to the PC.  Individual pulses vary greatly, but MPs are composed of 
a few, comparatively `smooth' subpulses which are much broader than those 
seen in the PC.  Conversely, the elements of PC emission have durations about 
equal to the sampling time and appear similar in character to the PC emission 
in B1822--09 (see Gil \etal\ 1994: Fig. 5) and the emission of B0656+14 (Weltevrede 
\etal\ 2006b: see Fig. 4).  Clearly, the PC is weak on an individual-pulse basis and 
affected by noise fluctuations, but the noise cannot account for its different character.

During the `Q' mode, integrated profiles of the MP have almost no linear or 
circular polarization.  As demonstrated in Suleymanova \etal\ (1988), this 
results from nearly equal power contributions by the two orthogonal polarization 
modes (hereafter OPMs).  Individual pulses contain significant linear polarization, 
but when aggregated, the polarization disappears.  Accounting for the 90$\deg$ 
separation of the two OPMs, the MP has a prominent linear PPA traverse of 
about --3.0 $\deg/\deg$ longitude (in both emission modes).

The PC, by contast, is highly linearly polarized (85\% at the peak); there is clearly 
one very dominant OPM.  Most striking is that within the errors, the PPA traverse 
is flat: 0~$\deg/\deg$~longitude.  

At `B' mode onset, both components undergo drastic changes.  The main pulse 
exhibits its well known modulation features discussed throughout this series.  
Drifting subpulses appear so rapidly that we are able to determine the time of 
the modal switch down to a single pulse (or two).  One of the OPMs dominates, 
resulting in an average profile with significant linear polarization: about 10\% 
at `B' mode onset, increasing to 40-50\% by `B' mode cessation (see Paper V).
The PC, by contrast, shuts off almost completely during the `B' mode.  Because of 
its weakness, it is impossible to determine how quickly the PC emission drops off.

Despite its weakness during the `B' mode, a trace of the PC can still be detected 
through its linear polarization.  In integrations of several thousand pulses, we 
see nothing of the PC in total power, but enough $L$ remains to define its PPA 
(see Fig.~\ref{fig:B0943_mode_polarization}).  Over some 20$\deg$ of longitude 
where the PC was present during `Q' mode, we now see polarized `noise,' with 
the flat traverse characteristic of the PC.  Note the contrast with the other PPAs 
outside of the PC and the MP that are random, as is expected of actual noise.  
Interestingly, Fig.~\ref{fig:B0943_mode_polarization} suggests that the `B'-mode 
PC polarization is orthogonal to that of its `Q'-mode counterpart, arguing that it 
may be the SPM which is seen here.  

Finally, we emphasize that while the B0943+10 MP is well understood in conal 
terms (\eg, Rankin 1993; hereafter ET VI), this PC feature is aberrant.  That the 
sightline traverse is highly tangential is clear from three different perspectives 
(see Paper I):  the dimensions and frequency evolution of its average profiles; the 
properties of its subbeam carousel; and that its exceptionally steep RF spectrum 
is due to the fact that its emission occurs inside the sightline circle at frequencies 
higher than about 400 MHz.  For all these reasons, the PC appears to fall outside 
any reasonable explanation within the hollow-cone/core model.

\section*{V. Meterwave Study of Pulsar B1822--09}
\label{1822}
As we have outlined above, pulsar B1822--09's mode-associated PC component 
and IP\footnote{
Motivated by the presence of an IP in B1822--09 we have conducted a search 
for such a feature in B0943+10.  This search has proven unsuccessful, but given 
the weakness of B1822--09's IP, a similar interpulse would appear absent for 
even a slightly different sight-line traverse, and given the weakness of B0943+10, 
a dim interpulse could be washed out in the noise.} 
have attracted great interest and have prompted extensive and repeated 
study over the years.  Virtually all previous single-pulse analyses of B1822--09 
have been carried out at frequencies above 1 GHz and usually with the Effelsberg 
telescope (\eg, Fowler \& Wright 1982).  Nonetheless, its tripartite profile and PPA 
traverse have proven difficult to interpret geometrically, and no existing study has 
provided a fully satisfactory model.  

\begin{figure} 
\begin{center}
\includegraphics[width=81mm]{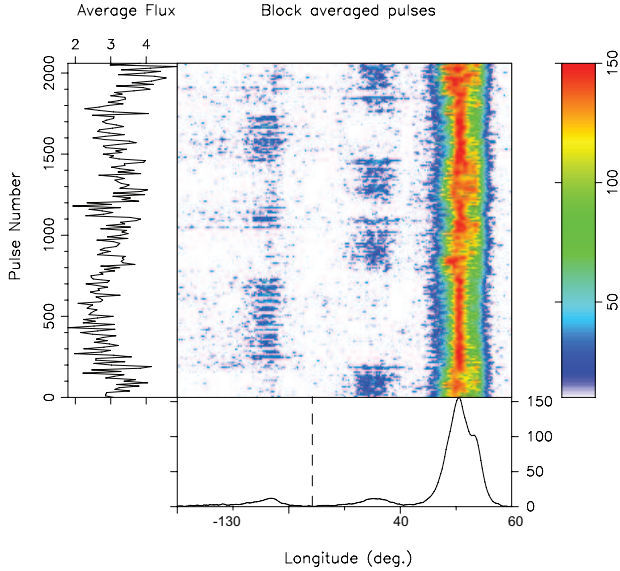}
\caption{The full 2077-pulse B1822--09 observation on MJD 53780 in 10-pulse 
averages.  The PC and MP components (on the right) are located at their actual 
relative longitudes; whereas the IP (on the left) is spliced into the plot prior to 
relative longitude +23\degr\ for convenience (exactly 140\degr\ of longitude is 
removed at this point).  The intensity scale at red saturates the MP (and is biased 
positively) in order to better show the IP and PC mode-changes.  The anti-correlation 
between the IP and PC is very clearly shown.  When one is `on,' the other is `off.'  
Note that the MP structure broadens in the `B' mode when the PC is present; 
whereas in the `Q' mode the 43-$P_1$ modulation is readily discernible in the IP.} 
\label{fig:1822_modes_diagram}
\end{center}
\end{figure}

Our interest in B1822--09 was prompted by its ostensible similarity to B0943+10.  
In order to explore this similarity fully, however, we find we need both to conduct 
some new analyses of the star's PSs and to interpret them in the context of an 
understanding of its emission geometry.  In particular, we have carried out the 
first in-depth analysis of B1822--09 at meter wavelengths, but even here we can 
make no easy assumptions about fully exploring the pulsar's effects because 
our two GMRT 325-MHz observations exhibit drastically different modal behavior.  
The MJD 53780 PS displays the star's characteristic mode-switching behavior: 
the two modes each endure for several minutes (around 200-500 pulses), with 
the overall profile being comprised fairly equally of both modes (see Figure \ref{fig:1822_modes_diagram}).  The MJD 54864 total-power PS, however, displays an 
hitherto unknown B1822--09 behavior: a 2106-length PS composed entirely 
of the `Q' mode, never once switching to the `B' mode.  Though much longer than 
usual, this `Q'-mode PS otherwise appears perfectly normal.

\begin{figure}
\begin{center}
{\large {\bf `B' mode}}
\includegraphics[width=80mm,angle=-90.]{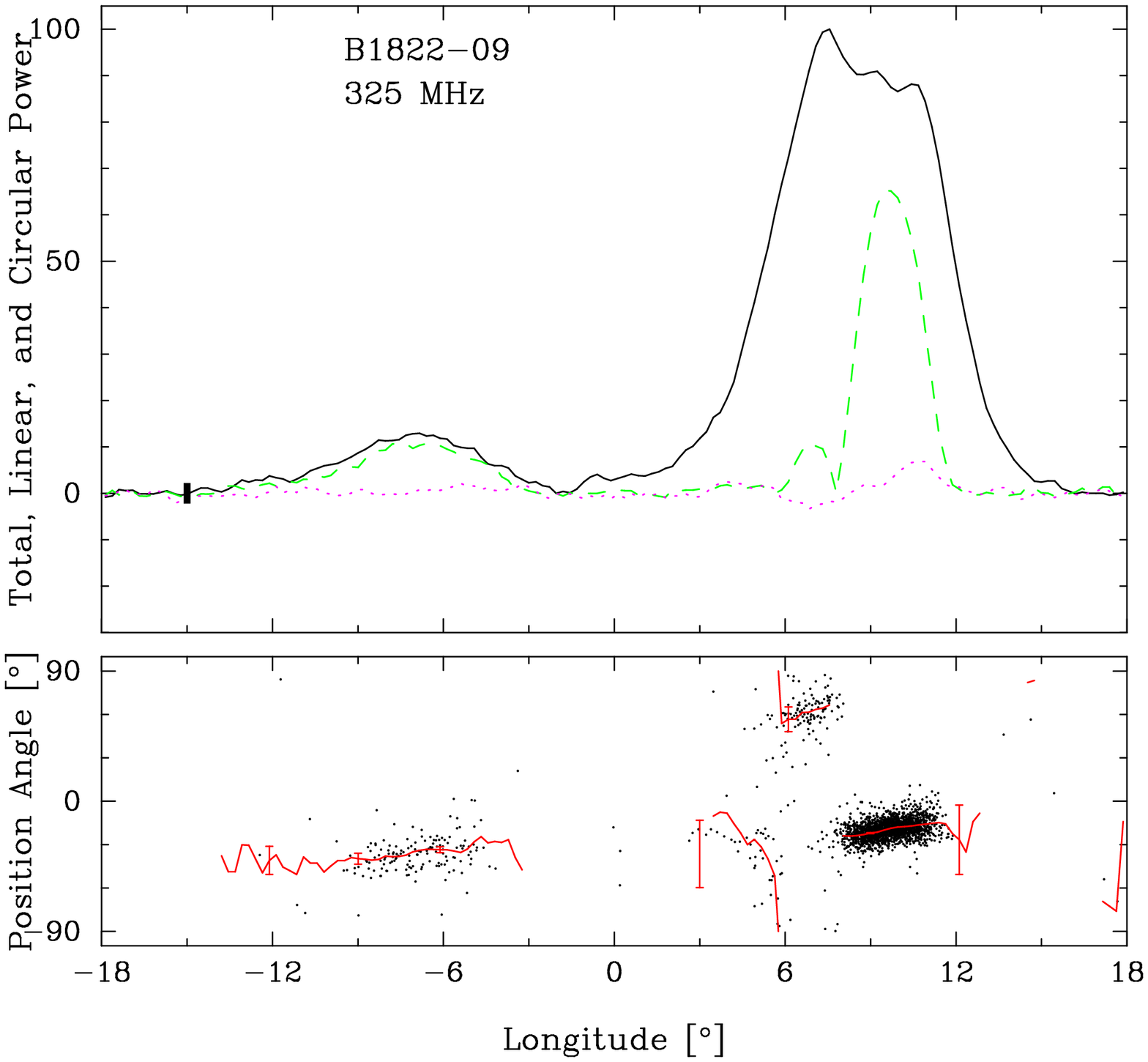}
{\large {\bf `Q' mode}}
\includegraphics[width=78mm,angle=-90.]{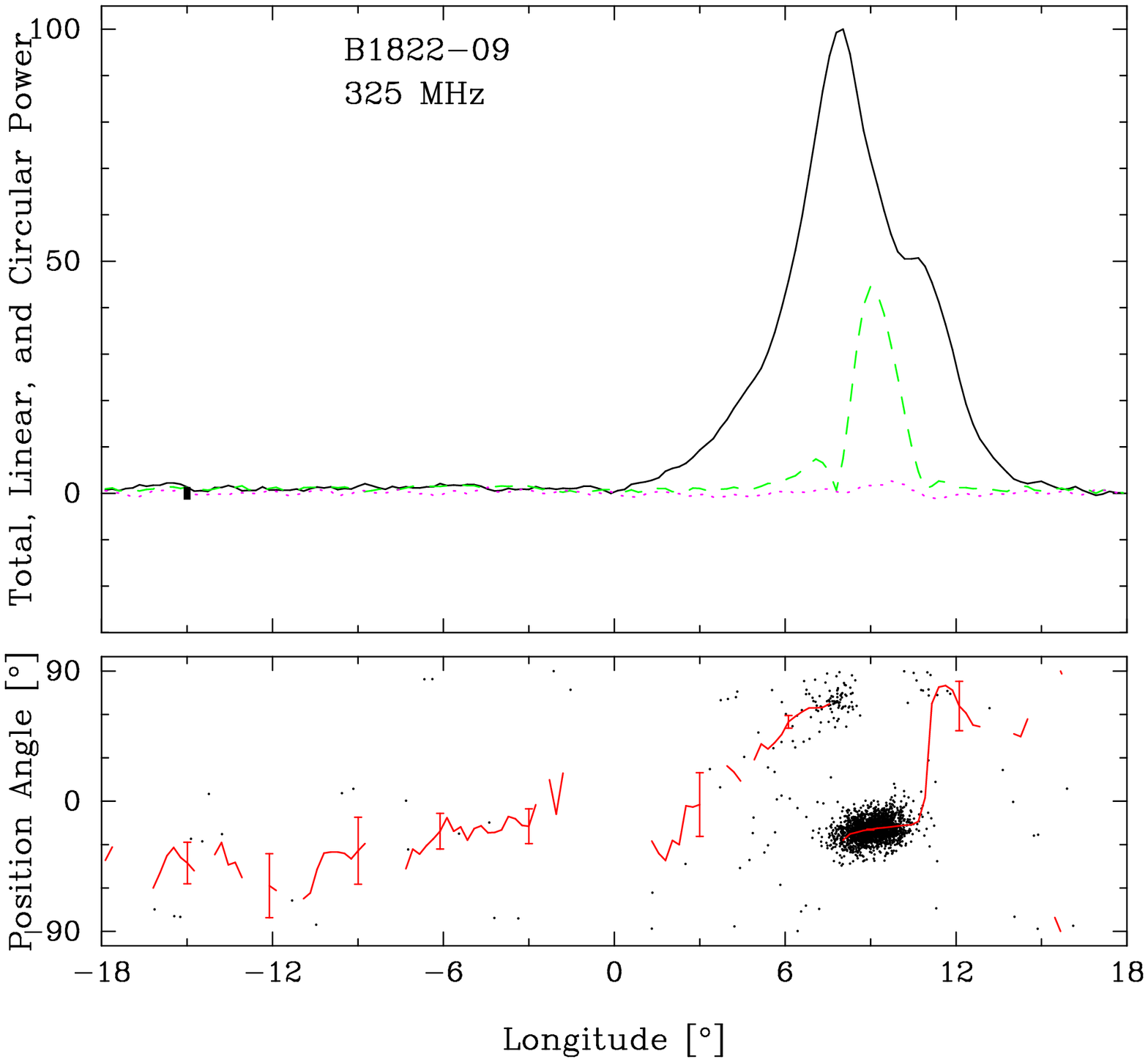}
\caption{Average profiles and polarization histograms of the PC and MP of 
B1822--09 during its `B' (pulse \#'s 1200-1455) and `Q' (\#'s 200-770) modes, 
in Fig.~\ref{fig:1822_modes_diagram}, respectively.  {\bf `B' mode (top)}: here 
the highly linearly polarized PC `turns on,'  and its nearly flat PPA traverse is 
remarkable.  Accounting for the 90$\deg$ OPM ``jump,'' the MP PPA is also 
essentially flat, suggesting a nearly central sightline traverse.  {\bf `Q' mode 
(bottom)}:  the conal components flanking the central MP feature have reduced 
intensity, so the profile is narrower.  PC emission is still faintly visible, along 
with a trace of its linear polarization.  The quantities plotted are the same as 
in Fig.~\ref{fig:B0943_mode_polarization}.} 
\label{fig:1822_Polarization}
\end{center}
\vspace{100mm}
\end{figure}

Figure~\ref{fig:1822_modes_diagram} shows the full 2077-pulse observation of 
MJD 53780 in 10-period averages; note that exactly 140\degr\ of longitude have 
been removed at +23\degr, so that all three emission features, the IP, PC and MP 
appear in this sequence.  The multiple modal transitions are obvious.  B1822--09's 
`Q' mode is characterized by the presence of its IP, along with a strong low frequency 
modulation feature.  During the `B' mode, the IP and the regular modulation cease 
almost completely, and a PC some 15$\deg$ before the main pulse `turns on'.  The 
figure also shows greater breadth and complexity in the MP during the `B' mode, 
that partially accounts for its greater aggregate intensity.  

\begin{figure} 
\begin{center}
\includegraphics[width=85mm]{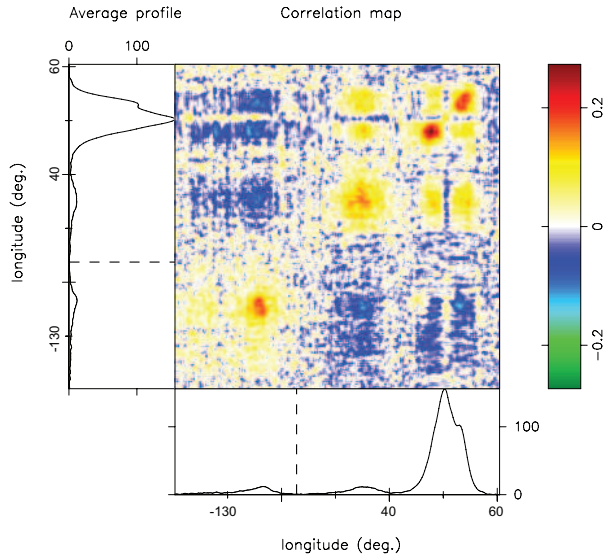}
\caption{Cross-correlation map of the MJD-53780 PS with itself at a delay of 
2$P_1$ calculated over the entire 2077 pulses, including both modes.  The 
main panel shows the correlation between different longitudes as marked by 
the profiles in both the side and bottom panels.  The leading and trailing edges 
of the MP are positively correlated with each other and the PC (and negatively
correlated with the IP), whereas the center of MP correlates with neither.  This 
strongly indicates that the MP has a three-zone emission structure, although 
this is not fully clear from its average profile.  The ordinate is delayed with respect 
to the abscissa, and the positive (negative) delay maps are shown below (above) 
the diagonal; these two maps are virtually identical, indicating correlations that 
are time-reversable.  The three-sigma error in the correlations is about 6\%.} 
\label{fig:lcorr_mixed_mode}
\end{center}
\end{figure}

Partial profiles corresponding to the two emission modes are given in Figure~\ref{fig:1822_Polarization}.    
The nearly complete linear polarization of the PC feature in the `B' mode (upper) 
is well known, but striking in contrast to that of the MP.  Note also that the PPA 
traverse of the PC is very flat, and that correlated PPAs at similar angles in the 
`Q'-mode profile (bottom) show that some PC power remains.  In fact, PPA rotation throughout the profile 
is very shallow: here we see only 
PPAs that are around --40 and +50\degr---presumably representing the two 
OPMs---and the same conclusion follows even for the largely depolarized IP 
[not shown, but see Gould \& Lyne (1998) at 1642 MHz].  Finally, the forms of 
the `B'- and `Q'-mode partial profiles are dramatically different:  Many total MP 
profiles show little structure, and care is needed in separating the emission modes to 
reveal the different contributions to MP power [\eg, see Gil \etal\ (1994): Fig. 1].  
Indeed, on the basis of the modal profiles here, we can only be sure that the MP 
{\em has} parts---that is, a bright central component as well as both a leading 
and trailing emission region.  Such evidence we already saw in 
Fig.~\ref{fig:1822_modes_diagram}, where fairly steady `B'-mode central-component 
power occurs together with leading and/or trailing emission.  

Finally, Figure~\ref{fig:lcorr_mixed_mode} shows a longitude-longitude correlation 
map for the entire 2077-pulse length of the MJD-53780 PS at a 2-$P_1$ delay.  
As we saw in Fig.~\ref{fig:1822_modes_diagram} this observation is comprised 
of about equal contributions of `B' and `Q'-mode intervals, so the map mixes the 
behaviours of the two modes.  Note, however, the strong correlations between the 
two sides of the MP and the other emission zones.  This is seen over all delays of 
a few pulses, and the nearly identical maps for negative and positive delays on 
either side of the diagonal are compatible with amplitude modulation.  The PC 
correlations with the sides of the MP reflect the greater MP activity in these regions 
during the `B' mode when the PC is present; the negative correlation with the IP, 
shows the opposite in the `Q' mode.  

 \subsection*{1. `B'urst mode in B1822--09}
Our observations provide only the four brief `B'-mode apparitions seen in the 
MJD-53780 observation of Fig.~\ref{fig:1822_modes_diagram}.  Fluctuation 
spectra of these intervals show no significant periodicities.  A weak `B'-mode 
modulation feature corresponding to about 11 $P_1$ has been reported at 
higher frequencies (\eg, Gil \etal\ 2004), but we find no evidence at all of such a 
modulation in our 325-MHz fluctuation spectra in any of the components. Also, 
a cross-correlation map similar to Fig.~\ref{fig:lcorr_mixed_mode} (not shown) 
for the `B`-mode interval of pulses 1200-1475 shows no significant correlation 
between the PC and MP emission regions.  

The `B' moniker in the literature derives from its greater MP intensity.  Much of this 
enhanced power owes to its broader profile, which in turn is due to its stronger 
leading and trailing components.  At `B'-mode onset, the pulsar's IP switches off 
almost completely, while a PC switches on.  The intensity of the PC is much less 
frequency dependent than that of either the IP or MP.  At high frequencies above 
1 GHz, the MP becomes progressively dimmer compared to the PC (Gil \etal\ 1994).  
Averaged over the four `B'-mode intervals, we find the PC 's peak intensity to be 
12\% that of the MP.  

The linear polarization of the PC is nearly complete and thus remarkably different 
from that of the MP (see Fig.~\ref{fig:1822_Polarization}).  This large $L$/$I$ 
extends across its entire width, such that nearly all of its power is in a single OPM.  
Its PPA traverse is linear and nearly flat with a slope somewhere between 0 and 
1.3$\deg/\deg$.   

The MP form and polarization structure is more typical of core/conal emission.  Its 
edges are completely depolarized, apparently by the usual OPM activity; whereas 
he middle of the MP (core?) shows a broad region of significant fractional linear 
that is divided by a 90\degr\ OPM-dominace ``jump''.  Overall, there is little 
rotation of the PPA under the MP:  the PPA under the leading part 
of the profile is essentially that of the well defined middle, and the ``jump'' at about 
+7\degr\ longitude is clearly OPM related.  With respect to $V$, there is a weak 
anti-symmetric signature that is centered at about +9\degr\ longitude, but it is not 
clear whether this is significant.  

We can now see clearly how it has been that B1822--09's profile is difficult to 
classify and interpret.  Little can be made of its ostensibly ``double'' average MP 
profile, and the modal partial profiles in Fig.~\ref{fig:1822_Polarization} are in turn 
quite complex.  We find unassailable evidence for a basic tripartite form---leading, 
middle and trailing---but even the modal profiles show us no simple triplicity.  That 
in the top panel of the above figure shows weak early and bright trailing emission 
around the central component, but other `B' episodes in Fig.~\ref{fig:1822_modes_diagram} 
have a reversed or more balanced character.  If then the central feature is of the 
core type, which seems a sensible premise on multiple grounds, then the MP's 
behavior is suggestive of the T or M profile class and a highly central 
sightline traverse.  

\begin{figure} 
\begin{center}
\includegraphics[width=78mm]{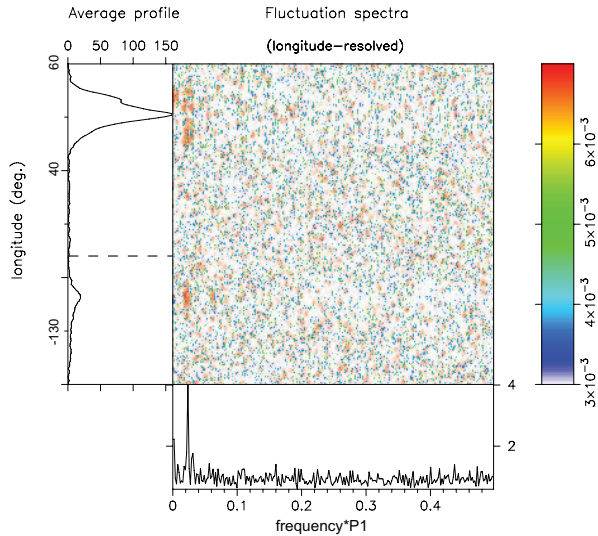}
\caption{Longitude-resolved fluctuation spectra of pulsar B1822--09's `Q' mode, 
averaged over pulses 221-750 of the MJD 53780 observation, using a 512-point 
FFT.  The MP is at the top of the left-hand panel, the PC in the center, and the 
IP at the bottom.  A strong feature at 0.023 c/$P_{1}$, corresponding to a $P_3$ 
of about 43 $P_1$, modulates both the MP and the IP, while the weak PC 
displays no discernible modulation.} 
\label{fig:1822_LRF}
\end{center}
\end{figure}

In this context, B1822--09's PC component is aberrant, in the sense that it has 
no clear interpretation within current understandings of the possibilities of polar 
cap emission.  Its flat PPA traverse and virtually complete linear polarization adds 
to this strangeness as does the character of its individual pulses.  Gil \etal's (1994) 
Fig. 5 plots a set of PC and MP single pulses with 50-$\mu$s sampling, and the 
difference between the respective two regions is startling:  one sees no subpulses 
in the PC as its emission elements typically have widths of only a single sample.  
The MP emission, by contrast shows emission structures that are several degrees 
wide---the subpulses with which we are familiar.  This ``spiky'' emission was also 
seen by Weltevrede \etal\ (2006b) in B0656+14, where they sometimes referred 
to its strikingly different character as ``rain''.  Also, we have seen above (see 
Fig.~\ref{fig:spiky_emission}) that the B0943+10 PC has the same characteristic.  

\subsection*{2. `Q'uiescent mode}
As we saw earlier in Fig.~\ref{fig:1822_modes_diagram}, the B1822--09 `Q' mode 
exhibits a strong and regular modulation affecting both its IP and MP.  Its period 
there can readily be estimated at about 40 $P_1$ (see also Weltevrede \etal\ 
2006a).  Figure~\ref{fig:1822_LRF} 
gives LRF spectra for the interval 221-750, and we see that the low frequency 
modulation produces a strong and narrow feature at 0.023 c/$P_1$.  The feature 
modulates both the MP and IP strongly and corresponds to a $P_3$ of some 43 
$P_1$.  An harmonic-resolved fluctuation spectrum (not shown) shows that the 
feature represents a mixture of amplitude and phase modulation.  Taking care 
to measure the primary period accurately by appropriately weighting two adjacent 
frequency components, we find a period of 43.75$\pm$1.0 $P_1$.  

\begin{figure} 
\begin{center}
\includegraphics[width=78mm]{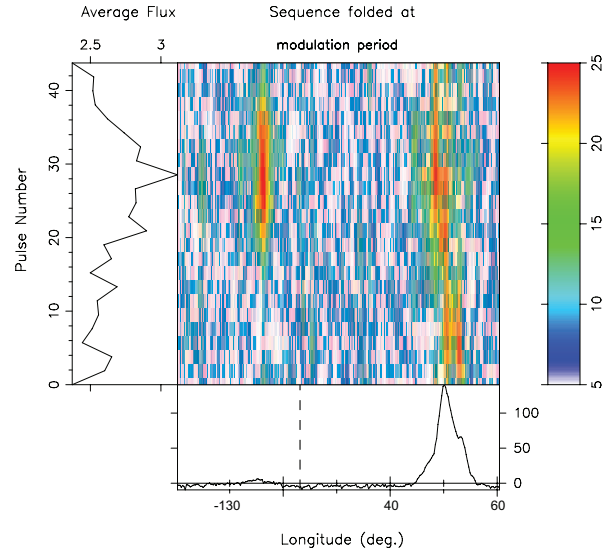}
\caption{`Q'-mode PS from Fig.~\ref{fig:1822_modes_diagram} folded at the primary 
modulation period.  Pulses 221-750 of the MJD 53780 observation are folded at 43.75 
$P_1$ corresponding to the bright modulation feature in Fig.~\ref{fig:1822_LRF}.  Here 
the unvarying `base' has been removed from the power in the central panel and the 
colour-scale compressed both at small and large intensities.  The modulation affects 
both the IP and MP (see the `base' profile in the bottom panel), producing primarily 
an amplitude (stationary) modulation in the IP and a phase modulation in the MP:  note 
the way in which the fluctuation power appears at only one phase in the IP; whereas 
in the MP, fluctuation power appears in both the leading and trailing regions of the 
profiles at different phases.} 
\label{fig:mod_fold}
\end{center}
\end{figure}

\begin{figure} 
\begin{center}
\includegraphics[width=78mm]{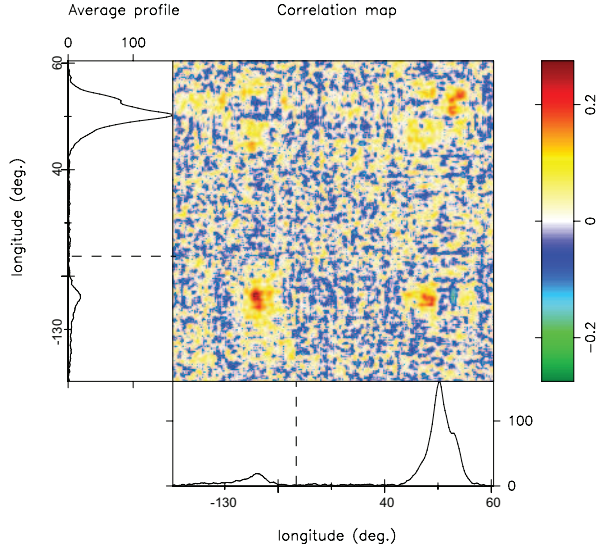}
\caption{Cross-correlations of the PS with itself at a delay of 2$P_1$, calculated 
over pulses 221-750 of the `Q' mode-only MJD 53780 observation.  Here we 
see a strong correlation of the IP with itself as well as a significant correlation of 
the delayed IP with the leading region of the MP profile, but not the reverse.  
Note also that it is a trailing region of the IP that is modulated, but a long leading 
region precedes it.  The 3-sigma error in the correlations is about 13\%.  See 
Fig.~\ref{fig:lcorr_mixed_mode} 
for details.} 
\label{fig:lcorr_Q_mode}
\end{center}
\end{figure}

The effects of this modulation periodicity can be further explored by folding the 
`Q'-mode PS at the 43.75-$P_1$ modulation period, and this display is shown 
in Figure~\ref{fig:mod_fold}.  Here the unfluctuating `base' power has been 
removed and the colour scale somewhat compressed to show the fluctuations 
more clearly.  The IP is fully modulated at this periodicity, so we see its power 
in only a particular region of the full cycle.  The MP, however, shows a ``wobble'' 
of fluctuation power extending from the leading to trailing regions of its overall 
profile.  Power in the leading profile region occurs nearly simultaneously with 
power in the IP, whereas the trailing MP region is bright at times when the IP 
power is at a minimum.  

\begin{figure} 
\begin{center}
\includegraphics[width=78mm]{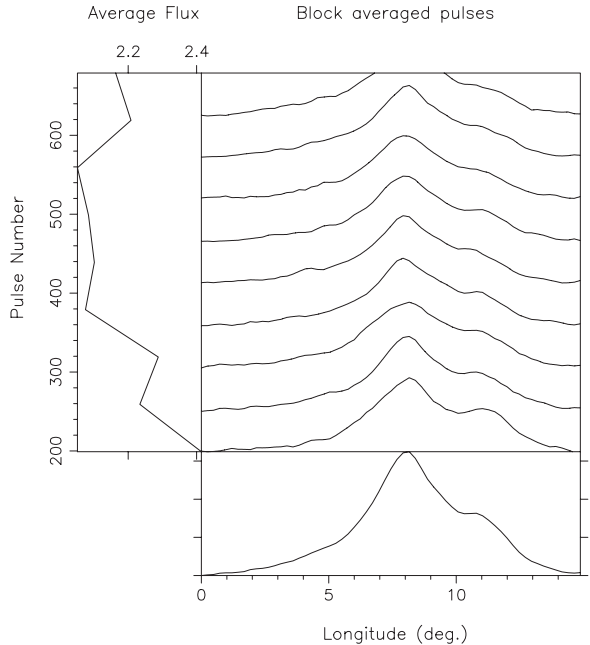}
\caption{Profile-shape changes in the MP of B1822--09 after `Q' mode onset, 
from the MJD 53780 observation.  Each profile is an average of 60 pulses.  
Directly after the `B'-to-`Q'-mode transition at about pulse 200, the profile has 
a prominent trailing component.  At later times after `Q'-mode onset, the 
leading profile region maintains a relatively stable intensity, while the trailing 
one gradually weakens.} 
\label{fig:pulse_shape_evolution}
\end{center}
\end{figure}

\begin{figure} 
\begin{center}
\includegraphics[height=78mm,angle=-90]{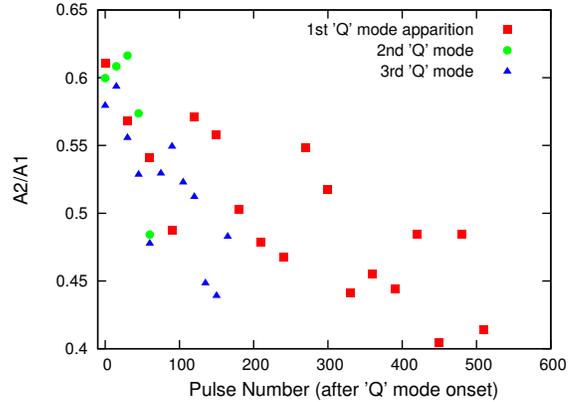}
\label{fig:A2_A1}
\caption{Changes in the trailing- ($A2$) to peak- ($A1$) amplitude ratio in the 
B1822--09 MP profile following `Q'-mode onset.  The amplitudes are measured 
from 30-pulse averages of the MJD 53780 observation.  $A1$ is measured at the 
peak of the profile and $A2$ as the intensity $\sim 3\deg$ later, at the peak of the 
trailing component.}
\label{fig:A2/A1_plot}
\end{center}
\end{figure}

Similarly, the longitude-longitude correlation map at a delay of 2 $P_1$ in 
Figure~\ref{fig:lcorr_Q_mode} shows significant correlation between the delayed 
IP and the leading regions of the MP; however, the map for the reverse (--2 $P_1$ 
delay) above the diagonal shows much less correlation.  This asymmetry is 
characteristic of a phase modulation that has a ``direction''.  Similar maps are 
obtained for other delays of a few periods.  Note also that only a trailing region 
of the IP is modulated, but a long weak region of emission precedes it.  

Finally, B1822--09's MP appears to exhibit secular changes over the several 
hundred pulses following `Q'-mode onset.  Figure~\ref{fig:pulse_shape_evolution} 
shows a set of 60-pulse averages following the first such onset in Fig.~\ref{fig:1822_modes_diagram}.  
Here we see that the power in the leading profile region remains fairly constant 
along with the intensity of the central component; whereas the power in the 
trailing profile region first exhibits a distinct component and thereafter declines 
progressively over the next 500 pulses.  That the three long `Q'-mode episodes in 
the MJD 53780 PS show a similar behavior is shown in Figure~\ref{fig:A2/A1_plot} 
where decreases of about 20\% relative intensity are seen over 200 pulses in all 
three cases.  Clearly, such a behaviour is very reminiscent of the changes seen 
in B0943+10 following its `B'-mode onsets, but on a very much shorter time scale.

\subsection*{3. The emission geometry of B1822--09's PC \& MP}
As we have seen above, B1822--09 presents a ``main pulse'' profile that has been 
very difficult to interpret.  First, it has not been clear whether the PC component was 
or was not a part of this ``main pulse'' region.  Indeed, it has been tempting to regard 
it as so, because the PC and MP are connected by a weak bridge of emission that 
would ostensibly seem to associate them.  Second, {in mixed average profiles of both 
the `B' and `Q' modes, the MP structure itself is not at 
all clear; some profiles show hardly more than a single component with 
a trailing ``bump'', and at best one can discern two barely resolved components.  

We now see clearly, however, that the PC is a completely different sort of ``animal'' 
than the MP:  it is comprised of a very unusual and distinct kind of emission elements, 
is highly linearly polarized, and it is modulated very differently from the MP.  It is truly 
and unmistakably a PC and not a part of the ``main pulse''.  In short, it is almost 
certainly not of a core/conal origin.  

Returning now to the MP, which indeed is the totality of the ``main pulse'', our various 
single pulse analyses have revealed that it is comprised of three very distinct regions, 
the leading, central and trailing regions.  The central region has a half-power width 
of some 3\degr\ and shows a very steady emission from pulse to pulse.  The leading 
and trailing regions, by contrast, are illuminated episodically and only occasionally 
at the same time---and in the `Q' mode their illumination is periodic with the same 
43-$P_1$ cycle as the IP.  The illumination of these leading and trailing regions is 
responsible in large part for the greater intensity of the `B' mode as is very clear from 
Fig.~\ref{fig:1822_modes_diagram}.  

For all these reasons, then, there can be very little doubt but that the MP of B1822--09 
should be classified as having a basically triple profile.  In some partial profiles, we 
see a suggestion of two conal rings in the leading or trailing regions, which would 
suggest a five-component {\bf M} profile, but such behaviour is not seen consistently 
enough to be certain.  Moreover, the softer spectrum of the central component, its 
regularity, lack of periodic modulation (and correlation with other profile features), 
and the hint of antisymmetric $V$ all suggest that this is a core component.  

We can then apply the quantitative geometrical methods of ET VI (Rankin 1993) to 
B1822--09's MP: Its PPA traverse is quite shallow, showing little orderly rotation---like 
that of B1237+25 (\eg, Srostlik \& Rankin 2005)---so we can take its central slope to 
be essentially exceedingly steep, indicating that the sightline passes almost exactly 
over the pulsar's magnetic pole.  The about 3\degr\ half-power width of this putative 
core further suggests a nearly orthogonal relationship between the star's rotation 
and magnetic axes as the angular width of the star's polar cap can be computed as 
2.8\degr.  Thus the magnetic latitude $\alpha$ and sightline circle $\zeta$ are both 
close to 90\degr.

With these constraints in mind, we can estimate what would be the total half-power 
angular sizes of the inner and outer conal regions and then compare them with 
the full width of the B1822--09 profile.  These respective conal widths are about 
9.5-10 (=4.33\degr$P_1^{-1/2}$) and 13\degr\ (=5.75\degr$P_1^{-1/2}$).  Referring 
conveniently to Fig.~\ref{fig:A2/A1_plot}, we can see immediately that the full outside 
width of the leading and trailing regions cannot be squared with 13\degr, but a width 
of 9.5-10\degr\ corresponding to an inner cone is fully plausible.  Therefore, we can 
conclude that B1822--09's MP is fully compatible with the inner-cone/core {\bf T} 
classification.  

\subsection*{4. The emission geometry of B1822--09's IP}
Our analyses also shed new light on the IP and its relationship to the MP and 
PC.  First, the IP is not a single symmetrical component, but rather a broad 
region of emission with a bright trailing component.  Figure~\ref{fig:MP_IP separation} 
gives a sensitive 325-MHz profile in which its somewhat double form and 
nearly 20\degr\ width are obvious.  In the older publications of Fowler \etal\ 
and Wright \& Fowler, one gets little sense of its extended form, and even in 
the Gil \etal\ work, the IP appears as a single asymmetric feature.  Clearly, 
the early observations lacked our sensitivity, and perhaps the IP changes 
its form at meter wavelengths, but in either case its broad and asymmetric 
character must be taken fully into account.  

\begin{figure} 
\begin{center}
\includegraphics[width=78mm]{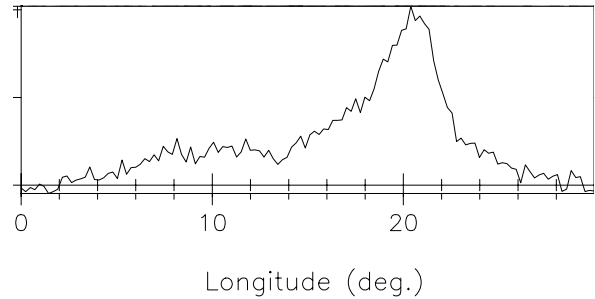}
\caption{A close-up average profile of the IP, averaged over all 2106 pulses of the 
MJD 54864 observation.  Dyks $\etal$ measure the MP-to-IP separation as $\sim$186$\deg$, 
but if instead of measuring from the peak or the center of the half-power point, we 
measure from some 6$\deg$ earlier---at the center of the IP---this separation 
reduces to almost precisely 180$\deg$.} 
\label{fig:MP_IP separation}
\end{center}
\end{figure}

The spacing of the IP from the MP is clearly shown in several of the previous 
figures, but the bottom panel of Fig.~\ref{fig:1822_modes_diagram} depicts 
their spacings with respect to the PC as well.  In this and the other diagrams, 
exactly 140\degr\ of longitude has been removed at longitude +23\degr, so 
the relationships between the three emission features can be measured 
conveniently.  Most obviously, we see that the interval between the IP and 
MP peaks is about 173\degr---that is, 33\degr\ as shown on the scale plus 
the removed 140\degr.  This is the measurement that most earlier workers 
have made, but Fig.~\ref{fig:MP_IP separation} above shows very clearly 
that the IP extends far on the leading side of the peak---so as to suggest a 
double profile form.  If the IP-MP spacing is instead measured from the IP 
``centroid''---some 6 or 7\degr\ earlier---then the resulting interval is very 
nearly 180\degr!  

But beyond these basically average-emission properties to consider in trying 
to understand the relationship between the IP and MP, we have seen above 
that there are also important dynamical connections.  First, both the IP and 
MP share the 43-$P_1$ modulation, and such modulation is usually conal in character.  
Second, the IP peak (including its entire trailing ``component'') shows strong 
positive correlation with the leading emission of the MP as well as negative 
correlation with its trailing region (as in the cross-correlation maps of Figs.~\ref{fig:lcorr_mixed_mode}~and~\ref{fig:lcorr_Q_mode} and the folded sequence in Fig.~\ref{fig:mod_fold}).  Given that these MP regions are conal in 
nature, it is tempting to conclude that the IP is comprised of a pair of conal 
components.   In short, the putative conal regions of the IP have an angular 
width comparable to that of the MP, and these respective regions behave 
similarly dynamically---so that their dynamic midpoints are again nearly 
180\degr\ apart.  

By contrast, the PC is an entirely different animal:  it is not opposite to the 
IP.  It shows a different type of emission.  It exhibits no periodic fluctuations.  
And it shows no structures that can be regarded as either conal or core-like.

\section*{VI. Implications for Current Models}
\label{discussion}
As has been demonstrated throughout this series of papers, the regular modulation 
features of B0943+10 are adequately understood in terms of the subbeam carousel 
model.  However, this is not the only extant model for B0943+10's remarkable 
subpulse-drift phenomena.  There are two papers, Clemens \& Rosen (2004) 
and Rosen \& Clemens (2008), that explored a non-radial oscillation model and 
then assessed whether it can produce the specific observations of Paper I.  They 
reanalyzed the 430-MHz PS of Paper I, confirmed these earlier results, and 
reiterated that the sideband feature occurs only within a small section of the full 
18-min observation.  They then suggest that the sidebands might be produced 
by a ``stochastic variation in pulse amplitudes,'' but clarify that if periodic amplitude 
modulation occurs within the drifting-subpulse sequences, then this would favor 
the carousel as opposed to the non-radial oscillation model.  We here present 
two further instances of the tertiary amplitude modulation, and instances in which 
the low frequency periodicity is primary are presented in Paper II and Paper IV.  
That these various instances are compatible with each other, exhibit orderly and 
very long secular variations and show complete frequency independence, would 
seem to favor the carousel model for drifting subpulses very strongly.  

We now turn our attention to the emission-reversal model proposed by Dyks \etal\ 
(2005) to explain the anti-correlation between the intensity of the PC and the IP 
in B1822--09.  They proposed that the PC and the IP are emitted from the same 
source which reverses emission direction during the different modes.  In their 
model, B1822--09 has a nearly orthogonal geometry (in agreement with our 
findings).  The physical source of the IP and the PC is located on the same pole as 
the MP, and the apparent IP results from inwardly directed emission from the 
source of the PC (Dyks \etal\ 2005: see Fig. 1).  This resolves the problem of 
information transfer between the poles:  if the IP and the PC are emitted from 
opposite poles, how can their behavior be so strongly anti-correlated?

However, we find that the MP and the IP are similar in their polarization properties 
and experience the same strong modulation in the `Q' mode.  Furthermore, they 
show a similar frequency dependence, whereas the PC is much less frequency dependent 
(Gil \etal\ 1994).  As argued above, 
they are both compatible with core-cone emission in the polar-cap model.  We also 
find that the IP and MP are separated by almost exactly 180$\deg$, not 186$\deg$ 
as found in Dyks \etal\ (2005).  This suggests that the MP and IP are both outwardly 
emitted and are produced by similar processes above the two respective magnetic 
poles.  The PC is markedly different and not compatible with core-cone emission, 
suggesting a physically different origin from the other two components.  

Another possibility for the emission-reversal model is that the source of the IP and 
PC is located on the opposite pole from the MP and that the IP is outwardly emitted, 
whereas the PC is inwardly emitted.  Propagation through the closed magnetosphere 
might explain the unique characteristics of the PC emission, and the MP-IP similarities
are easily explained in this model, but the question of information transfer between 
the poles arises again: the MP behavior is regulated by the same modes as the 
PC and IP. 

Thus, we conclude that the characteristics of B1822--09 are not easily interpreted 
with the emission reversal model as outlined above.  

\section*{VII. Summary \& Conclusions}
\label{conclusion}
The sideband features in pulsar B0943+10 were first seen in a 430-MHz PS 
discussed in Paper I.  Our analyses above have revealed a further two 
instances of B-mode sidebands in observations spanning more than 18 hours.  
Clearly, such tertiary modulation features, although remarkably rare, exhibit 
highly consistent characteristics.  We can then conclude with certainty that 
these sideband features indicate a physically significant periodicity, which 
within the subbeam-carousel model corresponds to the circulation time 
$\P3hat$.  

One instance above occurs soon after B-mode onset, while the other two 
follow it by about 90 mins.  In addition, evidence of a tertiary modulation in 
the form of a low frequency feature has been seen to occur several times in 
B-mode PSs at low frequencies (Papers II \& III)---and a single instance of a 
corresponding Q-mode feature was identified at 327 MHz (Paper IV).  All of 
these apparitions are consistent with a rotating-carousel subbeam system 
that has two discrete states: either the `beamlet' configuration {\bf is} sufficiently 
disordered so that no primary (``drift'') modulation is observed, or it is comprised 
of just 20 evenly spaced beamlets.

We also report the discovery of a bright precursor component in the Q-mode of 
B0943+10, falling some 50\degr\ longitude prior to the star's MP, which dims to 
nearly undetectable levels in the B-mode. This PC is almost fully linearly polarized 
with a nearly constant PPA traverse.  Its constituent radiation is ``spiky'' in character, 
as opposed to being comprised of broad subpulses---in this respect similar to that 
seen in pulsar B0656+14 (\eg, Weltevrede \etal\ 2006b).  Some residual PC 
emission is also seen in the B mode at a very low level.  In short, this PC feature 
appears to be neither conal nor core-like.  

These curious properties prompted us to compare B0943+10's modal emission 
characteristics with an ostensibly similar star, B1822--09, as it also exhibits two 
modes (also denoted B and Q) and a highly linearly polarized PC in one of its 
modes.  These pulsars have similar properties in terms of period, magnetic field, 
spindown energy and age.  However, their inferred emission geometries are 
very different:  B0943+10 has a small magnetic inclination angle and a highly 
tangential sightline traverse; whereas we argue above that B1822--09 has a 
nearly orthogonal magnetic geometry and that our sightline traverses its 
emission cone centrally.  Both stars have MP emission characteristics, both in 
terms of quantitative geometry and dynamics, that are fully comprehensible 
within a core-cone polar cap emission model.  In this context, we find that the 
PC emission is aberrant:  not only are its characteristics neither conal nor core-like, 
but the PCs' positions within the profiles of these two stars with well identified 
polar-cap emission regions would seem to rule out any similar origin.

The PC emission in both of these stars is nearly 100\% linearly polarized, much 
higher than what is typically seen in pulsar profiles.  With most pulsars, single 
pulses can be highly polarized (\eg, Mitra \etal\ 2009), but a number of depolarizing 
effects usually lead to substantial profile depolarisation.  Recently Johnston \& 
Weisberg (2006) have pointed out that young pulsars with higher spindown 
energies tend to show relatively simple and highly polarized average profiles. 
They hypothesize a possible time evolution for pulse profiles suggesting that high 
$\dot{E}$ pulsars have relatively simple profiles that arise from a single cone of 
emission high in the magnetosphere.  In turn, the depolarization effects are less 
effective at larger heights due to field line flaring, and thus the profiles retain their polarization over a 
wide range of frequencies.  This behavior seems compatible with B1822--09's 
PC polarization properties: it remains highly polarized from below 243 MHz 
(Gould \& Lyne 1998) to 3.1 GHz (Johnston \etal\ 2008).  

However, little more can be said; we thus far have little specific idea just where 
the PCs arise.  The PC longitude location is far from the MP, where the RVM predicted PPA 
is generally flat, as also seen in our observations.  However the full PPA traverse 
in these pulsars is not revealed, hence to see the effect of aberration/retardation 
at the PC location is difficult.  We note also that Petrova (2008a,b) has attempted 
to understand B1822--09's IP, PC and MP emission in terms of scattering effects 
within the pulsar magnetosphere.  

The discovery of a PC in B0943+10 which is so similar to the well-known PC of 
B1822--09 shows that such a component is not unique, and explaining its origins 
could have important ramifactions for our understanding of pulsar physics.  We 
should expect that there is one underlying physical process which regulates the 
modes and therefore the appearance of such diverse phenomena as subpulse 
drifting, polarization characterstics, pulse-shape dynamics, and the presence 
of a PC.  While the similarities between two stars are certainly telling, their 
dissimilarities are also important for any model purporting to explain the PC in 
either pulsar.  We cannot expect that the alignment of the axes, the sightline 
traverse, or the mode-dependent changes in brightness of the pulsar, 
play a role in the production of the PC and its modal behavior. 

%Maybe delete the above  paragraph.

\noindent {\bf Acknowledgments:}
We are pleased to acknowledge Jarek Dyks, Janusz Gil, Svetlana Suleymanova and 
Geoffrey Wright for their critical readings of the manuscript.  One of us (JMR) 
thanks the Anton Pannekoek Astronomical Institute for their generous hospitality 
and the NWO and ASTON for their Visitor Grants.  Portions of this work were 
carried out with support from US National Science Foundation Grants AST 
99-87654 and 08-07691.  Arecibo Observatory is operated by Cornell University 
under contract to the US NSF.  This work used the NASA ADS system.

\end{document}